# Using Grok to Avoid Personal Attacks While Correcting Misinformation on X

Kevin Matthe Caramancion
*Department of Defense Analysis*
*Naval Postgraduate School*, *U.S. Department of Defense (administered by the U.S. Navy)*
Monterey, California, United States
orcid.org/0000-0003-1332-3426

*Abstract*—Correcting misinformation in public online spaces often exposes users to hostility and ad hominem attacks, discouraging participation in corrective discourse. This study presents empirical evidence that invoking Grok, the native large language model on X, rather than directly confronting other users, is associated with different social responses during misinformation correction. Using an observational design, 100 correction replies across five high-conflict misinformation topics were analyzed, with corrections balanced between Grok-mediated and direct human-issued responses. The primary outcome was whether a correction received at least one ad hominem attack within a 24-hour window. Ad hominem attacks occurred in 72 percent of human-issued corrections and in none of the Grok-mediated corrections. A chi-square test confirmed a statistically significant association with a large effect size. These findings suggest that AI-mediated correction may alter the social dynamics of public disagreement by reducing interpersonal hostility during misinformation responses.

## I. Introduction

One persistent barrier in online misinformation correction is the personal risk that users perceive when challenging false claims. On platforms such as X, calling out inaccurate posts often triggers ad hominem attacks, hostility, and retaliatory behavior. This dynamic discourages users from participating in public fact-checking and weakens the collective capacity to counter deceptive narratives. Prior research has shown that online behavioral responses are shaped by perceived social threats, and that users often refrain from correcting misinformation when they anticipate personal backlash [1]. As online exchanges become increasingly contentious, understanding how individuals navigate these risks is critical to the broader study of misinformation control.

A growing behavioral pattern on X reveals a new strategy users employ to manage these risks. Many users now invoke Grok, the platform's native large language model, to generate corrective responses to misleading posts. Instead of issuing a correction personally, users prompt Grok to evaluate a claim and then present the AI-generated explanation. This approach allows users to distance themselves from the corrective act and shift responsibility for evaluating the falsehood to Grok.

Earlier findings suggest that users adjust their approaches to misinformation depending on their sense of safety and perceived vulnerability to personal attack, highlighting the social complexity underlying online correction behaviors [1]. In this context, invoking Grok may function as a protective buffer that enables users to challenge misinformation while reducing direct exposure to hostility. Despite extensive literature on misinformation detection, user vulnerability, and the psychological factors that shape online responses [1][2], there is minimal scholarship examining how AI-mediated tools are used as interpersonal shields in high-conflict settings.

The existing body of work focuses primarily on detection accuracy, cognitive susceptibility, exposure effects, or general behavioral patterns, but it has not examined how users deploy AI models as social intermediaries to mitigate personal risk. This gap limits our understanding of how AI-mediated communication reshapes the dynamics of misinformation correction in real-world environments.

This study addresses that gap by investigating how users on X invoke Grok when responding to misinformation and how such corrections are received socially. The analysis centers on two research questions:

*RQ1. How does the use of Grok for misinformation correction manifest across different misinformation topics on X?*

*RQ2. Is the likelihood of receiving personal or ad hominem attacks different for Grok-mediated corrections compared to direct human corrections during misinformation responses on X?*

The theoretical contribution of this paper lies in its integration of misinformation studies, online behavioral research, and AI-mediated interaction. It expands the current literature by identifying an understudied social function of AI systems in public discourse, specifically their role as intermediaries in conflict-prone correction contexts. The practical contribution offers insight into how users adapt their behavior in hostile online spaces and how platform-integrated AI features may shape patterns of correction and engagement.

Finally, this paper is organized as follows. The next section reviews relevant literature on misinformation correction, online conflict behavior, and AI-supported communication. The subsequent section details the data collection procedures, coding steps, and analytical approach used in this study. The findings and interpretations are then presented, followed by a discussion of implications and directions for future research.

## II. Literature Background

### A. The Social Risk of Online Correction, Hostility, and Ad Hominem Attacks

Research consistently shows that correcting misinformation in social settings exposes users to social and interpersonal risk. When individuals publicly challenge inaccurate content, they often face hostility or ad hominem responses, which discourages future participation in corrective behavior. Del Saz-Rubio (2023) found that impoliteness and aggressive language frequently emerge in contentious online exchanges, especially in political discourse [3]. In addition, Smith-Jones (2023) documented how individuals in online conversation use narrative and linguistic strategies to avoid direct interpersonal conflict when addressing hostile or sensitive topics, illustrating how mediated expressions can function as buffers in contentious exchanges [4]. These dynamics shape user expectations, making people more cautious or selective about when to correct misinformation. Ahn et al. (2015) demonstrated that mediated experiences can alter behavioral engagement by reducing perceived social and evaluative threat. Their findings show that when corrective or pro-social actions are framed through mediated systems rather than direct interpersonal confrontation, individuals report higher response efficacy and greater willingness to act. This supports the broader argument that social evaluation and anticipated negative feedback suppress public participation, while mediated or indirect engagement can lower these barriers and facilitate corrective behavior [5].

### B. Large Language Models and Misinformation: Risk, Opportunity, and Sociotechnical Consequences

Large language models play a complex role in the misinformation ecosystem. They can both amplify misinformation and support its correction. Chen and Shu (2024) reviewed current research and concluded that LLMs are powerful tools for misinformation detection while also recognizing their capacity to generate persuasive false content [6]. Pan et al. (2023) examined the dual nature of LLM influence and showed that these models can unintentionally assist misinformation campaigns when misused [7].

At the same time, models designed for evidence-based reasoning have shown promise in supporting correction. Zhou and colleagues (2024) demonstrated that claim verification assisted by LLMs can improve accuracy and consistency compared to unaided human judgments [8].

### C. LLMs as Intermediaries and Emerging AI-mediated Correction

An emerging body of research conceptualizes LLMs not only as informational tools but also as social intermediaries. Lim et al. (2025) described LLMs as communicative mediators that users may rely on when navigating conflict and misinformation in public platforms. This perspective suggests that LLMs can function as interpersonal buffers rather than merely classification systems [9].

Related work supports this trajectory. Gabriel et al. (2024) found that LLM-based interventions can shape the tone and direction of online conversations, indicating an interpersonal function beyond information processing [10]. In addition, Fu, Foell, Xu, and Hiniker (2024) show that people use AI mediated communication tools to craft or revise messages in high stakes interpersonal situations, often to manage tone, confidence, and potential conflict [11], which aligns with the idea of invoking Grok as a social buffer during misinformation corrections.

### D. Challenges Concerning Trust, Accuracy, and AI-mediated Communication

Despite their potential, concerns persist regarding the trustworthiness and reliability of LLM outputs. Mittelstadt et al. (2023) warned that LLMs often produce incorrect responses with persuasive confidence, raising risks when such systems are relied upon in factual disputes [12]. The Alan Turing Institute (2023) similarly cautioned that LLMs may unintentionally amplify harm in election-related misinformation contexts if used without careful evaluation [13].

These concerns underscore the importance of examining not only the accuracy of LLMs but also their social function when invoked by users. As Gabriel et al. (2024) note, AI-mediated communication can alter interpersonal dynamics in ways that traditional misinformation research has not yet fully addressed.

### E. The Literature Gap: The Missing Investigation of LLMs as Social Shields

The existing literature establishes that correcting misinformation is socially risky, that LLMs play an increasingly significant role in misinformation ecosystems, and that AI-mediated interactions may reshape online communication. However, no existing research empirically evaluates whether invoking an AI model is associated with reduced personal attacks during misinformation correction. Although conceptual discussions of LLMs as communicative intermediaries have emerged, empirical evidence remains limited. This study addresses that gap by examining whether Grok-mediated corrections are associated with different social responses compared to direct human corrections in high-conflict misinformation contexts on X.

## III. METHODOLOGY

### A. Study Design

This study employs an observational design that examines naturally occurring interactions on X through a structured sampling procedure. Rather than treating misinformation posts as the primary unit of analysis, the study treats individual correction replies as the unit of analysis. This approach aligns the sampling strategy directly with the research questions, which focus on how users experience social responses after issuing corrections, rather than on characteristics of the original misinformation posts themselves.

### B. Selection of Misinformation Claims

Five misinformation claims were selected based on prior verification by independent fact checking organizations. Claims were chosen to represent recurring and well documented false narratives that have circulated widely on X. These claims served only to define the topical scope of data collection and were not themselves treated as analytic units.

### C. Identification of Misinformation Posts

For each misinformation claim, multiple public posts on X asserting the claim as true were identified using keyword searches and platform search tools. Posts were eligible if they were publicly accessible, contained the target claim, and had at least one reply. No engagement threshold was imposed at this stage in order to avoid privileging viral or highly amplified content. These posts functioned solely as sources from which correction replies could be extracted.

### D. Extraction of Correction Replies

All replies to the eligible misinformation posts were manually reviewed to identify correction replies. A reply was classified as a correction if it explicitly disputed the accuracy of the original claim, presented factual counterinformation, or labeled the claim as false or misleading. Corrections were then categorized into two types. Direct corrections were those written entirely in the user's own voice without referencing AI systems. Grok mediated corrections were those that explicitly invoked Grok, referenced Grok's evaluation, or presented text clearly generated by Grok within the platform interface. This category also included replies that explicitly prompted Grok to assess the veracity of a claim, such as asking Grok whether a claim was true, provided that the prompt was clearly situated within a misinformation correction context and elicited or anticipated an evaluative response [10].

*E. Sampling Strategy and Reduction of Selection Bias*

To reduce selection bias associated with post visibility, reply ordering, and platform ranking algorithms, random sampling was conducted at the level of correction replies rather than misinformation posts. For each misinformation claim, all eligible correction replies were pooled by correction type. From each pool, ten direct corrections and ten Grok mediated corrections were randomly selected using a random number generator. This process yielded a total sample of 100 correction replies, consisting of 50 direct corrections and 50 Grok mediated corrections. By sampling corrections rather than posts, each eligible correction had an equal probability of inclusion regardless of the engagement level or prominence of the originating thread.

*F. Outcome Measurement and Coding Procedure*

The primary dependent variable was whether a correction reply received at least one ad hominem attack. For each sampled correction, all replies directed at that correction within a fixed observation window of 24 hours were collected. An ad hominem attack was defined as a response that targeted the correcting user personally rather than addressing the substance of the correction. Each correction was coded dichotomously, with a value of one indicating the presence of at least one ad hominem reply and a value of zero indicating none.

*G. Analytical Approach*

To address the first research question, descriptive analysis was used to document the presence of Grok-mediated corrections across multiple misinformation topics, rather than to estimate their prevalence in the broader platform. Because correction types were balanced by design, no population-level frequency claims were made.

To address the second research question, a chi-square test of independence was used to examine the association between correction type (Grok-mediated versus direct human correction) and the occurrence of ad hominem attacks. The dependent variable was coded dichotomously based on whether at least one ad hominem reply was observed within the defined observation window. Effect size was assessed using the phi coefficient. All analyses were conducted at the correction level.

*H. Definition of Ad Hominem Attack*

For the purposes of this study, an ad hominem attack was considered to have occurred when a reply directed at a correction targeted the correcting user personally rather than addressing the substance of the correction. This included insults, name calling, derogatory remarks, accusations about intelligence or character, and dismissive personal labels. Replies that criticized the argument, questioned the evidence, or expressed disagreement without personal targeting were not coded as ad hominem attacks. A correction was coded as having received an ad hominem attack if at least one such reply appeared within the defined observation window.

*I. Ethical Considerations*

All data were collected from publicly accessible posts on X. User identifiers were anonymized during analysis and reporting. No attempts were made to interact with users or influence platform behavior.

*J. Limitations*

This study is subject to several limitations. First, the analysis relies on publicly observable behavior and does not have access to users' underlying intentions or motivations. Although Grok mediated corrections were identified based on explicit references to Grok or recognizable AI generated output, it is not possible to determine whether users invoked Grok strategically to avoid personal attacks, out of convenience, or for other reasons. As a result, the findings reflect associations between observable correction behavior and social responses rather than direct evidence of user intent. Second, the study examines a limited number of misinformation claims and correction replies, which may constrain the generalizability of the results to other topics or platforms. Finally, because the data are observational and drawn from a single platform, causal conclusions about the effects of Grok invocation on social responses cannot be definitively established.

*K. Search Procedure*

Searches were conducted using X's search interface and web based search tools such as site restricted queries (for example, site:x.com). Searches were performed using a logged out browser session where possible to reduce personalization effects. The same search strings and time window were applied uniformly across all searches.

*L. Inclusion Criteria*

A correction reply was included if it met all of the following conditions:

1. It responded to a claim verified as false by an independent fact checking organization
2. It functioned as a corrective response to that claim
3. It was publicly accessible
4. It received at least one reply directed at it

*M. Exclusion Criteria*

A correction reply was excluded if:

1. It was clearly satirical or parody
2. It originated from automated or spam accounts
3. It referenced Grok without engaging in claim evaluation
4. It lacked a discernible misinformation context

IV. FINDINGS AND RESULTS

*A. Sample Overview*

The final dataset consisted of 100 misinformation correction replies collected from X across five misinformation topics: election denial, unsubstantiated vaccine claims, immigration and crime narratives, climate change denial, and claims that major mass shooting events were staged or fabricated. For each topic, 20 correction replies were analyzed, with equal representation of Grok-mediated corrections and direct human corrections. The resulting sample included 50 Grok-mediated corrections and 50 human-issued corrections.

*B. Descriptive Results*

Across all five misinformation topics, ad hominem attacks were observed in 36 of the 50 human-issued corrections,

corresponding to 72 percent of human corrections. In contrast, none of the 50 Grok-mediated corrections received ad hominem attacks. This pattern was consistent across all misinformation topics examined.

At the topic level, human-issued corrections received ad hominem responses in 8 of 10 cases for election denial claims, 6 of 10 cases for climate change denial claims, 9 of 10 cases for unsubstantiated vaccine claims, 6 of 10 cases for immigration and crime narratives, and 7 of 10 cases for mass shooting hoax claims. Grok-mediated corrections received zero ad hominem responses in all five topics.

*C. Inferential Analysis*

A chi-square test of independence was conducted to examine the association between correction type and the occurrence of ad hominem attacks. The analysis revealed a statistically significant association between correction type and ad hominem occurrence, $\chi^2(1, N = 100) = 56.25, p < .001$. The effect size was large, as indicated by a phi coefficient of .75. These results indicate that Grok-mediated corrections were associated with a substantially lower likelihood of receiving ad hominem attacks compared to direct human corrections.

*D. Tables and Figure Referenced*

Table 1 presents the distribution of ad hominem attacks by correction type across all misinformation topics.

TABLE I.  DISTRIBUTION OF AD HOMINEM ATTACKS BY CORRECTION TYPE (POOLED ACROSS TOPICS)

| Correction Type | Ad Hominem Yes | Ad Hominem No | Total |
|---|---|---|---|
| Human | 36 | 14 | 50 |
| Grok-mediated | 0 | 50 | 50 |
| Total | 36 | 64 | 100 |

*Note:* Each row represents a unique misinformation correction reply. Ad hominem was coded as the presence of at least one personal or character-based attack directed at the correcting user.

Table 2 presents a topic-level breakdown of ad hominem outcomes by correction type.

TABLE II.  AD HOMINEM OUTCOMES BY MISINFORMATION TOPIC AND CORRECTION TYPE

| Misinformation Topic | Human Yes | Human No | Grok Yes | Grok No |
|---|---|---|---|---|
| Election denial | 8 | 2 | 0 | 10 |
| Climate change denial | 6 | 4 | 0 | 10 |
| Unsubstantiated vaccine claims | 9 | 1 | 0 | 10 |
| Immigration and crime narratives | 6 | 4 | 0 | 10 |
| Mass shooting hoax claims | 7 | 3 | 0 | 10 |

*Note:* Each topic includes 20 correction replies, evenly split between human-issued and Grok-mediated corrections.

Figure 1 shows the percentage of misinformation corrections that received at least one ad hominem response, comparing direct human-issued corrections with Grok-mediated corrections. Across all five misinformation topics, 72 percent of human-issued corrections received ad hominem attacks, whereas none of the Grok-mediated corrections received ad hominem responses.

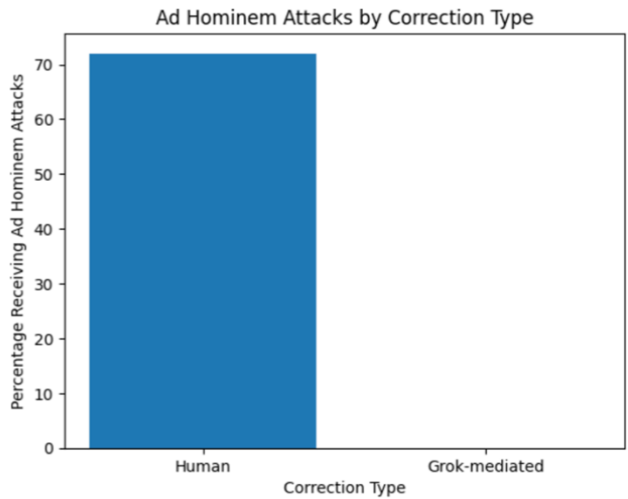

Fig. 1. illustrates the percentage of corrections receiving at least one ad hominem response by correction type

## V. ANALYSES AND INTERPRETATIONS

The findings directly address the second research question by demonstrating a strong association between correction type and the occurrence of ad hominem attacks during misinformation correction on X. Across all five misinformation topics examined, direct human-issued corrections were frequently met with personal attacks, whereas Grok-mediated corrections received no ad hominem responses in the observed sample. This pattern was consistent across politically, scientifically, and socially contentious domains, suggesting that the association is not confined to a single issue area or narrative context.

One plausible interpretation of this pattern is that invoking Grok alters the perceived social dynamics of misinformation correction. When users present corrective information through Grok, the act of correction may be reframed from a personal challenge into a mediated evaluation attributed to an external system. This shift may reduce perceptions of interpersonal confrontation, thereby lowering the likelihood that other users respond with personal hostility. In contrast, direct human corrections more clearly signal interpersonal disagreement, which prior research has shown can escalate conflict and provoke ad hominem responses in online environments.

The results further suggest that Grok may function as a social intermediary rather than merely an informational aid. By attributing corrective judgments to an AI system, users appear to redistribute responsibility for the correction itself, potentially deflecting personal blame or antagonism. This interpretation is consistent with broader research on online interaction showing that depersonalization, authority attribution, and mediated communication can influence how audiences respond to contested claims.

With respect to the first research question, the analysis confirms that Grok-mediated corrections are observable across multiple misinformation domains, including election denial, public health misinformation, immigration narratives, climate change denial, and claims that mass shooting events were staged. However, because correction types were balanced by design, these findings do not estimate the prevalence of Grok use on the platform, nor do they imply that

Grok-mediated corrections are representative of typical user behavior on X.

At the same time, these findings should not be interpreted as evidence that Grok inherently prevents hostility or that users deliberately invoke Grok to avoid personal attacks. Because the study relies on observational data, it cannot determine users' motivations for invoking Grok or isolate the causal mechanisms underlying the observed association. Differences in tone, phrasing, perceived legitimacy, or audience characteristics may also contribute to the absence of ad hominem responses observed in Grok-mediated corrections.

Finally, the consistency of the observed pattern across diverse misinformation topics suggests that the dynamics captured in this study may reflect broader features of online conflict rather than topic-specific sensitivities. If AI-mediated corrections systematically reduce personal attacks across domains, this raises important questions about how platform-integrated AI tools reshape public discourse, influence user participation, and affect individuals' willingness to engage in misinformation correction despite the risk of harassment.

## VI. Conclusion and Future Work

Correcting misinformation in public online spaces is not merely a cognitive task. It is a social act carried out under conditions of visibility, judgment, and risk. This study examined how users on X navigate that risk by invoking Grok during misinformation corrections and how such corrections are received compared to direct human responses. Across five high conflict misinformation domains, Grok-mediated corrections were consistently associated with the absence of ad hominem attacks, while direct human corrections were frequently met with personal hostility.

These findings suggest that the value of AI systems in misinformation contexts may extend beyond accuracy or efficiency. When users attribute corrective judgments to a platform-integrated AI, the correction itself appears to shift from an interpersonal confrontation to a mediated evaluation. In doing so, AI systems such as Grok may subtly reshape the social dynamics of public disagreement by redistributing responsibility, reducing perceived threat, and altering how corrective messages are interpreted by others. Importantly, this study does not claim that Grok prevents hostility or that users consciously deploy AI tools to shield themselves from attack. Rather, it demonstrates a robust association between AI-mediated correction and markedly different social responses.

The implications of this association are consequential. Public misinformation correction relies on individuals' willingness to speak up despite the risk of backlash. If AI-mediated correction lowers that risk, even indirectly, it may influence who participates in corrective discourse and how often. At the same time, reliance on AI systems introduces new concerns related to trust, accountability, and the authority granted to algorithmic judgments. Understanding these trade-offs requires moving beyond questions of whether AI systems are accurate and toward questions of how they function socially within contested environments.

Future research should expand this line of inquiry in several directions. Larger-scale and longitudinal studies could assess whether the observed patterns persist across time, platforms, and cultural contexts. Qualitative analyses could examine how tone, attribution, and perceived legitimacy interact with AI-mediated correction to shape audience responses. Experimental and survey-based approaches could explore users' motivations for invoking AI systems and whether perceived social safety influences correction behavior. Finally, comparative studies across different AI models and platform designs could clarify whether the effects observed here are specific to Grok or reflect broader dynamics of AI-mediated communication.

As platforms continue to integrate AI systems into everyday interaction, the question is no longer only whether these tools can help identify falsehoods. It is whether they can change the social conditions under which truth is contested. By showing that AI-mediated corrections are associated with substantially different interpersonal outcomes, this study highlights the need to treat AI not only as an informational resource but also as a participant in the social ecology of online discourse.

## Data Availability